# On the Origin of Time

Ernst Karl Kunst

**Herein it is shown that mass and time in the rest frame as well as relativistically enlarged mass and dilated time in the moving frame are of like origin. This implies that the former are generated by the movement of a fourth spatial dimension of matter relative to a four-dimensional manifold.**

In the previous work on quantization of velocity, length and time it has been proven that among others Einstein's "relativity of simultaneity" [1] rests on a misinterpretation of the principle of relativity and the correct interpretation of the Lorentz transformation to predict an expansion dx' = dx$\gamma_0$ of the dimension (dx) parallel to the velocity-vector of moving bodies - where $\gamma_0$ is the Lorentz factor based on quantized velocity $v_0$ - rather than the so called FitzGerald-Lorentz contraction [2]. Furthermore, this relativistic expansion of dx has been shown to imply that ρ' = ρ - density of mass in the moving frame and in the rest frame, respectively - and, thus, to be the cause of the relativistic rise of energy or mass $m_t$ = dt'$_x v_0$/c = dt$_x \gamma_0 v_0$/c, where $m_t$ is mass due to time dilation and dt$_x$ = dx/c. In connection with the definition of action as the product of energy and time the result also has been derived that a smallest or fundamental length $\lambda_0$ = √h and a quantum of time $\tau_0$ = √h/c exists and that apart from a numerical factor in the case of the hydrogen (H-) atom must be valid $\lambda_0$/c = $\tau_0$ = m, where m is rest mass of the atom.

The following investigates whether those results in connection with relativistic time dilation allow any conclusion as to the very cause of time in the rest frame.

Suppose a H-atom moving inertially at velocity $v_0$ relative to an identical atom at rest so that

$$\gamma_0 = \frac{E'}{E} = \frac{m'}{m} = \frac{dx'}{dx} = \frac{dt'}{dt}$$

is valid, where E means rest energy, dx and dt the geometrical dimension parallel to the velocity vector and time in the rest frame, respectively and the dashed values denote the respective ones of the moving atom. If dt$_x$ = dx/c the relativistic rise of mass can be written as

$$\frac{m' v_0}{c} = \frac{1}{c^2}\sqrt{E'^2 - E^2} = \frac{dx' v_0}{c}\rho\, dy\, dz = \left(\frac{dt_x v_0}{c} + \frac{dt_x v_0}{c}(\gamma_0 - 1)\right) c\rho\, dy\, dz, \qquad (1)$$

whereas the dilation of time attains the form

$$\frac{dt' v_0}{c} = \sqrt{dt'^2 - dt^2} = \frac{dt v_0}{c} + \frac{dt v_0}{c}(\gamma_0 - 1).  \tag{2}$$

Clearly (2) coincides with (1) - apart from the constant (invariant) factor "$c\rho dy dz$" in (1). This implies relativistic mass and time to be equivalent and generated in the volume $V' = dx'dy'dz' = dx'dydz$ by the movement of the latter at velocity $v_0$ in the (arbitrary) x-direction of three-dimensional space - as has been shown before in [2].

On the other hand, according to the principle of relativity for any observer based at the dashed system (considered moving), the rest mass of the H-atom and the quantum of time would be $m' = \lambda_0'/c = \tau_0'$. Thus, as observed from the resting system, according to (1) $\tau_0'$ must be composite:

$$m' = \tau'_0 = \sqrt{\tau_0^2 + \left(\left(\frac{dt_x v_0}{c} + \frac{dt_x v_0}{c}(\gamma_0 - 1)\right) c\rho\, dy\, dz\right)^2}. \tag{3}$$

This implies also rest mass and relativistic mass and, therewith, time in the rest frame and dilated time, respectively, to be equivalent that means of like origin. In other words: analogously to the generation of relativistic mass and dilated time, rest mass and time in the rest frame must be generated by the movement of a - for any observer - hidden spatial dimension of the H-atom. This leaves the only conclusion that $\lambda_0$ is a fourth geometrical dimension of the H-atom and physical time (quantum of time) in the rest frame as well as rest mass to result from the motion of $\lambda_0$ - the latter being orthogonal to three-dimensional space - relative to a four-dimensional manifold.

Given that the fundamental length in $R^1$ is $\lambda_1$, then the hypotenuse of the smallest possible Pythagorean triangle in the respective manifold is altogether the fundamental length in the latter, namely

$$\sqrt{2} \times \lambda_1 \text{ in } R^2,$$
$$\sqrt{3} \times \lambda_1 \text{ in } R^3,$$
$$2 \times \lambda_1 \text{ in } R^4.$$

In the previous work [2] for $\lambda_0$ the value $\lambda_0 = mc = \sqrt{h}$ has been derived, ignoring its four-dimensional nature. But from the foregoing is clear that the fundamental length in $R^4$ is $2\lambda_1$ and that this fact has to be taken into consideration yet. Hence, it must be valid

$$2\lambda_1 = \lambda_0 = \sqrt{h} \tag{4}$$

and for the quantum of time

$$\tau_1 = \frac{\lambda_1}{c} = \frac{\lambda_0}{2c} = \frac{\sqrt{h}}{2c}. \tag{5}$$

Thus, the real value of the quantum of time derived from the fundamental length in $R^4$ is only half the one previously predicted in [2] so that the ratio

$$\frac{\overline{T}}{\tau_0} = \frac{h}{\tau_0 \Gamma} = n, \ (n = 1, 2, ...) \tag{6}$$

of the mean life-times of short-lived particle resonances and of the time quant given there has to be corrected to

$$\frac{\overline{T}}{\tau_1} = \frac{2\overline{T}}{\tau_0} = \frac{2h}{\tau_0 \Gamma} = 2 \times n (n = 1, 2, ...), \tag{7}$$

where T means life-time and $\Gamma$ full width. This implies that all life-times in units of the time quant computed (in [2]) according to (6) must be doubled in accordance with (7). Thus, all ratios (6), which delivered (nearly) integers plus a half, become - after doubling - integers now and the life-times of the top quark ($\Gamma \approx 1.55$ GeV) and of the 1370 MeV meson ($\Gamma \approx 385$ MeV) found at Brookhaven [3] are $2\tau_1$ and $8\tau_1$ (instead of $\tau_0$ and $4\tau_0$), respectively.

The result that all ratios (7) are (nearly) full integers strongly supports the conclusion as to the fourth-dimensional nature of rest mass and time. If the theory is correct it follows, time to be linked to matter, and to space only to the extent as the latter possesses mass or energy, furthermore, that no particle resonances with life-times < $\tau_1$ = 1.357628 × $10^{-24}$ s in nature exist.